\documentstyle[prl,preprint,aps]{revtex}
\begin{document}
\draft
\tighten
\preprint{ }
\title{Limit on T-Violating P-Conserving $\rho N N$ Interaction \\
from the  $\gamma$ Decay of $^{57}$Fe}
\author{M.T. Ressell}
\address{W.K. Kellogg Radiation Laboratory, 106-38, California
Institute of Technology, Pasadena, CA 91125}
\author{J. Engel}
\address{Department of Physics and Astronomy,
University of North Carolina, Chapel Hill, NC 27599}
\author{P. Vogel}
\address{Department of Physics, 161-33, California Institute
of Technology, Pasadena, CA 91125}
\date{\today}
\maketitle
\begin{abstract}

We use the experimental limit on the interference of M1 and E2
multipoles in
the $\gamma$-decay of $^{57}$Fe to bound the time-reversal-violating
parity-conserving $\rho N N$ vertex.  Our approach is a large-basis
shell-model calculation of the interference.  We find an upper limit
on the
parameter $\bar{g}_{\rho}$, the relative strength of the T-violating
$\rho N N$ vertex, of close to $10^{-2}$, a value similar to the
best limits
from other kinds of experiments.

\end{abstract}

\narrowtext

For many years it has been difficult to compare the
quality of limits on
time-reversal-violating parity-conserving (TVPC)
interactions coming from
different low-energy experiments.  The experiments
typically limit
observables unique to themselves, and before
comparisons can be made,  these
limits must be translated into a common TVPC quantity.  It turns
out that a
convenient measure of nuclear TVPC interactions is
the dimensionless ratio,
often called $\bar{g}_{\rho}$\cite{r:Haxton}, of
the TVPC $\rho$-meson --
nucleon coupling to the normal strong coupling $g_{\rho}$.
Among the other
mesons only those with axial-vector couplings can
transmit TVPC interactions
between nucleons via a single exchange\cite{r:Simonius},
and they are
significantly heavier than the $\rho$ and consequently
less effective in
nuclei.  It is therefore reasonable to treat all
TVPC nucleon-nucleon
interactions as arising from $\rho$ exchange, and to
use $\bar{g}_{\rho}$ to
parameterize their strength.

Experimental upper limits on several quantities,
including the electric
dipole moments of the neutron and of $^{199}$Hg\cite{r:Haxton},
and a
correlation in the scattering of polarized neutrons from aligned
$^{165}$Ho\cite{r:Engel}, have been translated into limits on
$\bar{g}_{\rho}$, constraining it to be less than about $10^{-2}$.
A number
of other experiments, looking e.g.  for the violation of detailed
balance\cite{r:DB}, remain to be similarly interpreted.  In this
paper we
report an examination of a 1977 experiment\cite{r:Boehm} that
searched for
interference between M1 and E2 radiation in the
$\gamma$-decay of the first
$5/2^-$ state in $^{57}$Fe to the first $3/2^-$ state.
(Neither is the
ground state; the two have excitation energies of 137 keV
and 14 keV.)  Our
approach was to diagonalize the strong nuclear hamiltonian in the
shell-model, and then treat the TVPC $\rho$-exchange
interaction as a
perturbation that causes the interference by mixing higher-lying
states into
the two involved in the transition.  Ref.\ \cite{r:Beyer} employed this
method to constrain the TVPC coupling of the $A_1$ meson to the nucleon
from
the same experiment, but used what we argue is too small a
model space.  In
addition, the lighter and more commonly considered $\rho$ meson was
neglected completely.

The M1-E2 interference that signals T violation can be
expressed in terms of
$\sin\eta$,\footnote{$\sin\eta$ is directly proportional to
the experimental
correlation $({\bf J}\!\cdot\! {\bf q}\!\times\! {\bf
E}) ({\bf J}\!\cdot\!{\bf q})({\bf J}\!\cdot\!{\bf E})$,
where {\bf J} is the
quantization axis of the initial nucleus, {\bf E} is
the photon electric
field vector, and {\bf q} is the photon direction\cite{r:Boehm}.}
the
imaginary, part of the multipole mixing ratio $\delta$\cite{r:Henley},
which is defined
as\cite{r:Bied}
\begin{equation}
\delta = \frac
{\left\langle\,{J_f}\,\left\|\,T^{M_1}\,
            \right\|\,{J_i}\,\right\rangle}
{\left\langle\,{J_f}\,\left\|\,T^{E_2}\,
            \right\|\,{J_i}\,\right\rangle}
=|\delta|(\cos\eta+i\sin\eta).
\end{equation}
In Ref. \cite{r:Boehm} the upper limit on $|\sin\eta|$ was
expressed in the
form of a measured value that included zero within
experimental accuracy:
\begin{equation}
|\sin\eta| = (3.1\pm 6.5)\times 10^{-4}.
\label{sinexp}
\end{equation}
The contributions to $\eta$ can be written as
\begin{equation}
\eta =
\varepsilon_{E2}-\varepsilon_{M1}+\xi
\hspace{1cm}
\varepsilon_{E2},\varepsilon_{M1}, \xi \ll 1
\end{equation}
where the last term $\xi$ is represents effects of final
state interactions,
which have been shown\cite{r:DKV} to be smaller than the
upper limit in Eq.\ (\ref{sinexp}).

In first-order perturbation theory, the difference between the two
$\varepsilon$'s is\cite{r:Blin}
\begin{eqnarray}
i(\varepsilon_{E2} -\varepsilon_{M1}) = &\nonumber \\
&\sum_n
\frac {
{\left\langle{J_f}\left|{V_\rho}
            \right|{nJ_f}\right\rangle}}{E_f-E_n}
\left(\frac{\left\langle{nJ_f}\left\|{E2}
            \right\|{J_i}\right\rangle}
{\left\langle{J_f}\left\|{E2}
            \right\|{J_i}\right\rangle}
-\frac{\left\langle{nJ_f}\left\|{M1}
            \right\|{J_i}\right\rangle}
{\left\langle{J_f}\left\|{M1}
            \right\|{J_i}\right\rangle}\right)\nonumber\\
+&\sum_n
\frac {
{\left\langle{nJ_i}\left|{V_\rho}
            \right|{J_i}\right\rangle}}{E_i-E_n}
\left(\frac{\left\langle{J_f}\left\|{E2}
            \right\|{nJ_i}\right\rangle}
{\left\langle{J_f}\left\|{E2}
            \right\|{J_i}\right\rangle}
-\frac{\left\langle{J_f}\left\|{M1}
            \right\|{nJ_i}\right\rangle}
{\left\langle{J_f}\left\|{M1}
            \right\|{J_i}\right\rangle}\right) ~.
\label{eps}
\end{eqnarray}
With nucleons represented by $i,j$, the two-body $\rho$-exchange
potential has the form
\begin{eqnarray}
\label{e:exchange}
 V^{\rho} & =& \sum_{i,j}{\cal V}^{\rho}_{i,j} ~ [\tau_i
\times \tau_j ]_3
\\
{\cal V}^{\rho}_{i,j} & = &{m_{\rho}^3   g_{\rho}^2
\bar{g}_{\rho} ~ \mu_v \over 4 \pi M^2} ~ {e^{-m_{\rho}r_{ij}} \over
m_{\rho}^3 r_{ij}^3 } (1 + m_{\rho} r_{ij} ) (\mbox{\boldmath
$\sigma$}_i -
\mbox{\boldmath $\sigma$}_j ) \cdot {\bf l} ~ , \nonumber
\end{eqnarray}
where ${\bf r}_{ij} = {\bf r}_i - {\bf r}_j$, ${\bf l} =
{\bf r}_{ij} \times
\frac{1}{2} ({\bf p}_i - {\bf p}_j)$, $\mu_v = 3.70$ n.m.\ is
the isovector
nucleon magnetic moment, $M$ is the nucleon mass, $g_{\rho}=2.79$
is the
normal strong $\rho NN$ coupling, and $\bar{g}_{\rho}$ is the
dimensionless ratio of the TVPC coupling to $g_{\rho}$ that we
are trying to
constrain.  After choosing a model space and interaction (and
a reasonable
prescription for treating short-range
correlations\cite{r:Haxton,r:Beyer}),
we can use this formalism in a shell-model calculation to
translate the
experimental limit on $\sin \eta$ to a limit on $\bar{g}_{\rho}$.

The issues surrounding the calculation are more complicated than they
initially appear, however.  To evaluate the phases in
Eq.\ (\ref{eps}) one
needs, in principle, the wave functions and energies of
$all$ $3/2^-$ and
$5/2^-$ states in $^{57}$Fe.  To obtain them, one ought to
diagonalize the best available nuclear hamiltonian for 17
valence nucleons
moving freely in the $pf$ shell.  Such a space has an m-scheme
dimension of $\sim 4.5 \times 10^8$.  At the other extreme is a
minimal model space, based on the well-established shell
closure at $N$ or
$Z$ = 28, consisting of 3 valence neutrons in the
$(2p_{3/2}, 1f_{5/2},
2p_{1/2})$ shells and the remaining 14 nucleons in the
$1f_{7/2}$ subshell.
This ``small space" is the one used in Ref.\ \cite{r:Beyer}
and contains few
enough states to allow direct diagonalization of any Hamiltonian.
Unfortunately this space artificially restricts the M1 strength
from any
given state because it doesn't allow the important
$1f_{7/2}$-$1f_{5/2}$
spin-flip transition.  Consequently, in the calculations
described here we used a ``large space",
constructed by allowing a single proton or neutron to move out of the
$1f_{7/2}$ shell into any one of the other subshells.  The large space
contains 23604 m-scheme states, forcing an approximate diagonalization.

To obtain the approximate wave functions we used the
Lanczos algorithm as
implemented in the shell-model code CRUNCHER \cite{r:Resler} and its
auxiliary codes, with slight modifications to accommodate
the imaginary
two-body matrix elements of the interaction in Eq.\ (\ref{e:exchange}).
Since it was not practical to calculate all $J^{\pi} =
{3\over{2}}^-$ and
${5\over{2}}^-$ wave functions (there are 2052 and 2755 of these,
respectively), we adopted a procedure expounded in Ref.\
\cite{r:Bloom} to
obtain Gamow-Teller strength functions.  We first used
the Lanczos algorithm
to obtain the lowest ${3\over{2}}^-$ and ${5\over{2}}^-$
states in $^{57}$Fe
to high precision.  Next we created a ``collective" E2 or M1
state by acting
on the parent state with the relevant operator.  We then
used the collective
state as the initial basis vector for an approximate Lanczos-based
diagonalization of higher-lying states, yielding
pseudo-eigenvectors (PSEVs),
which approximate the true states.  We typically performed about 100
Lanczos
iterations, resulting in about 100 PSEVs for each of the $J_i$'s.
In the
Lanczos approach the lowest (and highest) several PSEVs
are quite accurate
representations of the corresponding eigenstates,
while at intermediate
energies the PSEV's converge more slowly, and after $\approx 100$
iterations
each still has contributions from tens to hundreds of actual
eigenstates.  It is easy to see, however, that all of the strength is
contained in
these PSEVs, which we used for the states $nJ_i$ and $nJ_f$ in Eq.\
(\ref{eps}).  The wave functions used in
each of the four terms in Eq.\ (\ref{eps}) were slightly different
since they originated from different initial collective states.

So far we have not mentioned our choice of interaction.
There are several
effective interactions on the market, but (unfortunately)
we did not know
which was the best in this space.  We were able to
test the sensitivity of
our results to the choice of Hamiltonian, however, and so
used 3 different
$pf$-shell interactions:  the FPVH interaction of
\cite{r:vanHees}, the TBLC8
interaction of \cite{r:vanderMerwe}, and the FPBPN
interaction (the FPD6
interaction of \cite{r:Richter} with the single particle
energies modified to
fit $^{56}$Ni \cite{r:Brown}).  Each of these interactions
reproduced the
energy spectrum of low lying states in $^{57}$Fe reasonably
well.  The spread
in the calculated values of the phase
$\varepsilon_{E2} -\varepsilon_{M1}$
with these interactions provided a rough measure of
theoretical uncertainty.

The last component of the calculation was the choice of
effective E2 and M1
operators for each force.  The matrix element $\langle {5\over{2}}
\vert\vert$ E2 or M1 $\vert\vert {3\over{2}} \rangle$ normalizes
each term in
Eq.\ (\ref{eps}).  Since the M1 matrix element (in the
denominator in the second and fourth term in Eq.\ (\ref{eps})) is
very small,
it is particularly important, and we chose effective $g$-values for
the M1
operator in order to reproduce it accurately.  Our prescription
was to fix
all of the M1 $g$-values, except for the isoscalar spin
piece ($g_{IS}^s$),
at their free nucleon values.  For each interaction we then chose
$g_{IS}^s$
to give the correct matrix element for the first transition.
The sign of the
matrix element is not known, so we chose it consistently amongst
the forces to obtain the most reasonable values for the
set of $g_{IS}^s$'s.

For the E2 operator a similar procedure gave unrealistic values for
the
effective charges $e_p$ and $e_n$; we therefore adopted
the ``canonical''
values $e_p = 1.5 e$ and $e_n = 0.5 e$ for all of the
interactions.  These
values result in reasonable agreement with the first E2 matrix element,
especially for the FPBPN force.  In addition, the
final phase $\epsilon_{E2}$
is only weakly dependent on the choice of the $E2$ effective
charges.  Table
I summarizes the E2 and M1 matrix elements and total strengths
for the few
lowest states in both the large and small spaces.
(The TBLC8 force shares a
common heritage with the FPVH force and, since the
results are similar, we
omit TBLC8 from the tables).  In the large space
the total strength for
both multipoles is
relatively insensitive to the force chosen.  However, the M1
strength is about a factor of 10
larger than in the small space, dramatically
illustrating the importance of
including the $1f_{5/2}$ level.

How much did the non-convergence of the intermediate
PSEVs affect the
results?  The answer is very little for the E2 part
of the phase, because
the strength is concentrated at low energies and the
energy denominator in Eq.\ (\ref{eps}) enhances
the contribution of the low
lying converged states and reduces the effects
of the higher lying states.
In Figure 1 we show the distribution of E2 strength
for the ${3\over{2}}^- \rightarrow n{5\over{2}}^-$
transitions (dashed line).
It is completely dominated by transitions among the
converged states.  A
similar result holds for the E2 in the ${5\over{2}}^- \rightarrow
n{3\over{2}}^-$ direction.  Though the effects of the energy
denominator are also at work in M1 piece of
the phase, the distribution of M1 strength complicates matters.
In Figure 2
we show the total M1 strength for the ${3\over{2}}^- \rightarrow
n{5\over{2}}^-$ transitions (dotted line); a broad resonance
is visible at
$\sim 12$ MeV.  Although this is the region where the
PSEVs are unconverged, the M1 part of the phase nonetheless
seems to be
represented reasonably well.  We make this statement
after varying the  number
of Lanczos iterations and hence the number of PSEVs (converged and
unconverged) to see if the phase changed appreciably
as the approximations
became more accurate.  The size of the dependence is
illustrated in Table II,
where the E2 and M1 parts of the phase $\eta$
(with $\bar{g}_{\rho} = 1$) are  listed for several numbers
of iterations and for two different interactions.  The E2 phases show
essentially no dependence on the number of iterations
(as implied above) and
the M1 phases are not affected dramatically, indicating that
the true result
is not far from our best approximation.

The results in Table II allow us to constrain the parameter
$\bar{g}_{\rho}$
and estimate the uncertainty.  The FPVH and the FPBPN forces give very
similar results for each piece of the phase and the final phases are
very close.  The TBCL8 interaction gives a similar result,
$\varepsilon_{E2} -\varepsilon_{M1} = -24.2 \times 10^{-3}$.
The lack of dependence on the interaction suggests that
the uncertainty in the
results is not large.
Table II also suggests that the phase is insensitive to the
size of the model
space, but this turns out to be a coincidence.  In the small
space, all of
the M1 piece of the phase lies at very low excitation energy,
mirroring the
initial downward peak at 2--3 MeV in Figure 2.
But the fall in the phase from 3--10 MeV and the subsequent rise due
to the M1 resonance are not present in the small space
and so the agreement
on the final value of $\varepsilon_{M1}$ between
the two model spaces is accidental.

The entries in Table I were evaluated with
$\bar{g}_{\rho}$ = 1.  Neglecting
theoretical error, which we have argued should be fairly
small, and averaging
the results from the FPVH and FPBPN forces in the
large space, we conclude
that $\vert \varepsilon_{E2} -\varepsilon_{M1}
\vert / \bar{g}_{\rho} = 16.4
\times 10^{-3}$.  The experimental value for $|\sin\eta|$, Eq.
(\ref{sinexp}), then implies that \begin{equation}
|\bar{g}_{\rho}| = (2
\pm 4) \times 10^{-2}. \label{limit} \end{equation}
This number is comparable
to the best limits from other experiments.  Limits on electric dipole
moments, for example, correspond to $|\bar{g}_{\rho}|~
\raisebox{-.25ex}{$ \stackrel{<}{\scriptstyle
\sim}$}\ 10^{-2}$, and the new
data on neutron-holmium\cite{r:Gould} scattering
yields $|\bar{g}_{\rho}| =
(2.3 \pm 2.1) 10^{-2}$.  Perhaps coincidentally,
all these very different
experiments give roughly the same limit.
It has been suggested\cite{r:Orlov}, however, that upcoming
detailed balance experiments, which go through complicated compound
nuclear states,
may provide limits that are better than these
by two orders of magnitude.
Even though recent theoretical work\cite{r:Framp,r:Khrip}
indicates that one cannot expect $\bar{g}_{\rho}$ to be
much larger than
$10^{-8}$, it remains worthwhile to translate limits from other
experiments into limits on $\bar{g}_{\rho}$.  Theoretical
expectations are
easily and often confounded, and it's important to know
which of the many experiments reported in the
literature (and still to come) have the best chance of
actually seeing time reversal violation.

We thank David Resler for several useful discussions and help with
the modifications to the code CRUNCHER.
This work was supported in part by the U.S. Department of Energy under
Grants DE-FG05-94ER40827 and DE-FG03-88ER-40397 and
by the U.S. National
Science Foundation under Grants PHY94-12818 and PHY94-20470.
M.T.R. is supported by the Weingart Foundation. Part of
the computing was carried out at Lawrence Livermore
National Laboratory,
operated under the auspices of the U.S.\ Department of
Energy under grant W-7405-ENG-48.

\begin{figure}
\caption[Figure 1] {The piece of $\varepsilon_{E2}$ arising from
E2 transitions with $J_f = {3\over{2}} \leftarrow  nJ_i = {5\over{2}}$
using the FPBPN force in the large space.
The solid line is the sum from eq. (\ref{eps}).  The
points correspond to the individual points in the sum.
The dashed line is the individual B(E2)
in $e^2$fm$^4$ divided by a factor
of $10^5$.  It is apparent that  $\varepsilon_{E2}$ is well converged
at low excitation energies.
}
\label{fig1}
\end{figure}

\begin{figure}
\caption[Figure 2] {The piece of $\varepsilon_{M1}$ arising from
M1 transitions with $J_f = {3\over{2}}
\leftarrow  nJ_i = {5\over{2}}$
using the FPBPN force in the large space.
The solid line is the sum from eq. (\ref{eps}).  The
points correspond to the individual points in the sum.
The dashed line is the individual B(M1) in nuclear
magnetons divided by a
factor
of $100$.  $\varepsilon_{M1}$ is well converged
at excitation energies above 20 MeV.
}
\label{fig2}
\end{figure}

\begin{table}
\caption[Table I]{The absolute values of the E2 and M1
matrix elements for the FPVH and FPBPN forces compared to the
experimentally determined values.
The total calculated E2 and M1 strengths are also included.
}
\begin{tabular}{ccccc}
Transition & Experiment & FPVH(small space) &
FPVH(large space) & FPBPN(large space) \\
 & & & & \\
\tableline
 & & & & \\
M1 ${5\over{2}})_1 \rightarrow {3\over{2}})_1$
& 0.113 & 0.113 & 0.113 &.0.113 \\
E2 ${5\over{2}})_1 \rightarrow {3\over{2}})_1$
& 13.35 & 7.091 & 6.866 & 12.88 \\
M1 ${5\over{2}})_2 \rightarrow {3\over{2}})_1$
& 0.344 & 1.202 & 0.806 & 0.858 \\
E2 ${5\over{2}})_2 \rightarrow {3\over{2}})_1$
& 27.83 & 27.62 & 36.38 & 37.73 \\
M1 ${3\over{2}})_2 \rightarrow {5\over{2}})_1$
& 0.298 & 0.141 & 0.018 & 0.126 \\
E2 ${3\over{2}})_2 \rightarrow {5\over{2}})_1$
& 3.208 & 11.66 & 14.12 & 13.87 \\
 & & & & \\
\tableline
 & & & & \\
B(M1) ${3\over{2}})_1 \rightarrow n{5\over{2}}$ &
& 1.154 & 11.56 & 10.57 \\
B(E2) ${3\over{2}})_1 \rightarrow n{5\over{2}}$ &
& 211.7 & 440.7 & 480.5 \\
B(M1) ${5\over{2}})_1 \rightarrow n{3\over{2}}$ &
& 0.598 & 5.18 & 3.63 \\
B(E2) ${5\over{2}})_1 \rightarrow n{3\over{2}}$ &
& 43.8 & 97.0 & 101.5. \\
\end{tabular}
\label{table1}
\end{table}

\begin{table}
\caption[Table II]{The phases, with $\bar{g}_{\rho} = 1$ in Eq.\
(\ref{e:exchange}) and multiplied by a factor of
$10^3$, computed with the FPVH and FPBPN interactions. The number of
Lanczos
iterations is
listed to illustrate the convergence of the phases.  200 iterations
were performed only for the cases listed.  Each of columns
3--6 corresponds
to one of the
terms in Eq.\ (\ref{eps}); for example, the heading
$\varepsilon_{E2} (n3/2)$
corresponds to the first term of the equation with
$J_i = {5\over{2}}$ and $nJ_f = {3\over{2}}$.  The last column
contains the final phase calculated according to Eq.\ (\ref{eps}).}
\begin{tabular}{ccccccc}
Force(Space) & Lanczos & $\varepsilon_{E2} (n3/2)$ &
$\varepsilon_{E2} (n5/2)$
 & $\varepsilon_{M1} (n3/2)$
& $\varepsilon_{M1} (n5/2)$ & $\varepsilon_{E2} -\varepsilon_{M1}$ \\
 & Iterations & & & & & \\
 & & & & & & \\
\tableline
 & & & & & \\
 FPVH(small) & Complete & -5.08 & -5.15 & 7.04 & 10.10 & -27.6 \\
 FPBPN(small) & Complete & -6.50 & -5.84 & 7.94 & 13.03 & -33.3 \\
 & & & & & \\
\tableline
 & & & & & \\
 FPVH(large) & 100 & -3.239 & -2.322 & -0.814 & 12.681 & -17.4 \\
 FPVH(large) & 200 & -3.239 & & -0.691 & & \\
 FPBPN(large) & 60 & -2.436 & -2.306 & 1.913 & 8.5052 & -15.2 \\
 FPBPN(large) & 100 & -2.425 & -2.298 & 2.135 & 8.4813 & -15.3 \\
 FPBPN(large) & 200 & & & 1.769 & & \\
\end{tabular}
\label{table2}
\end{table}

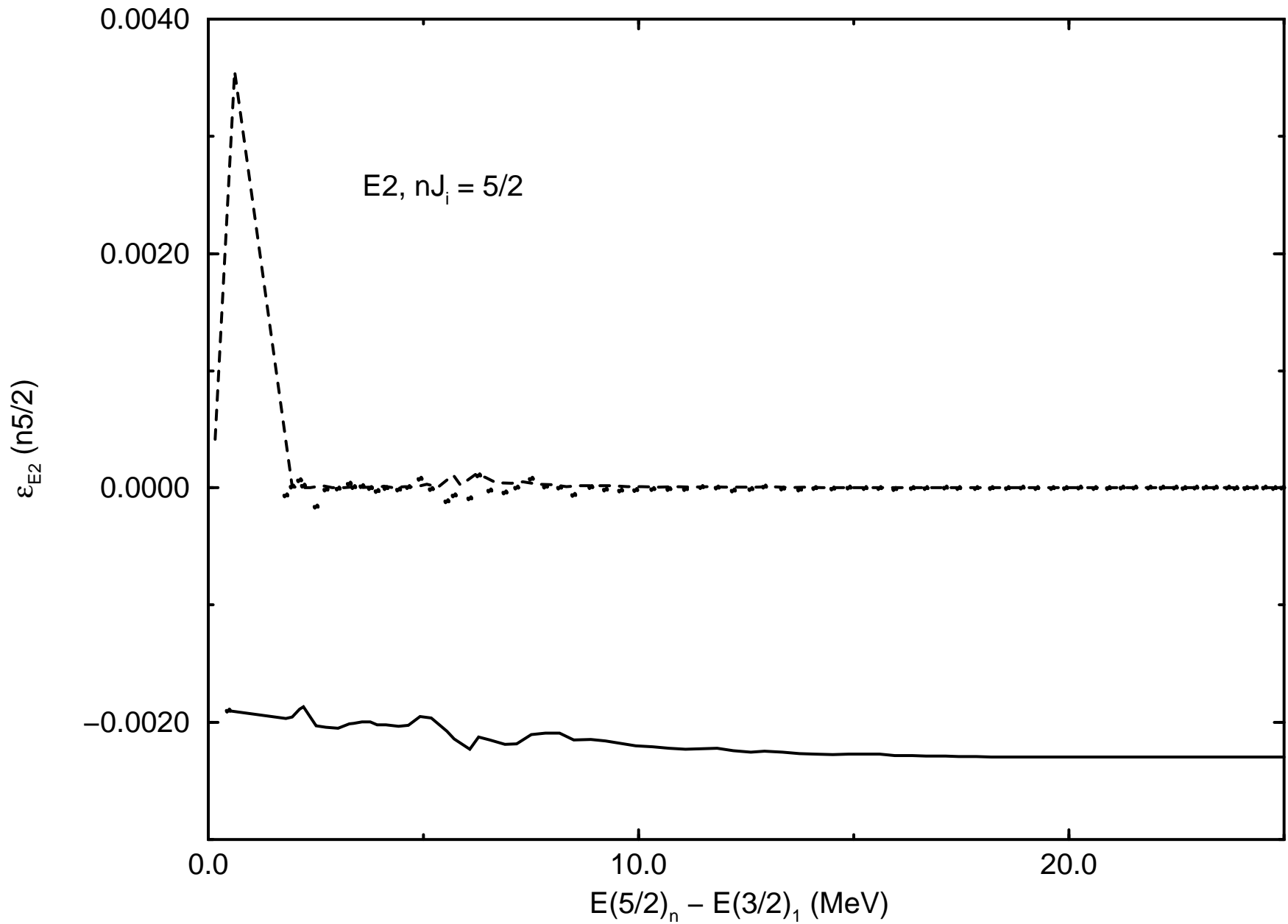
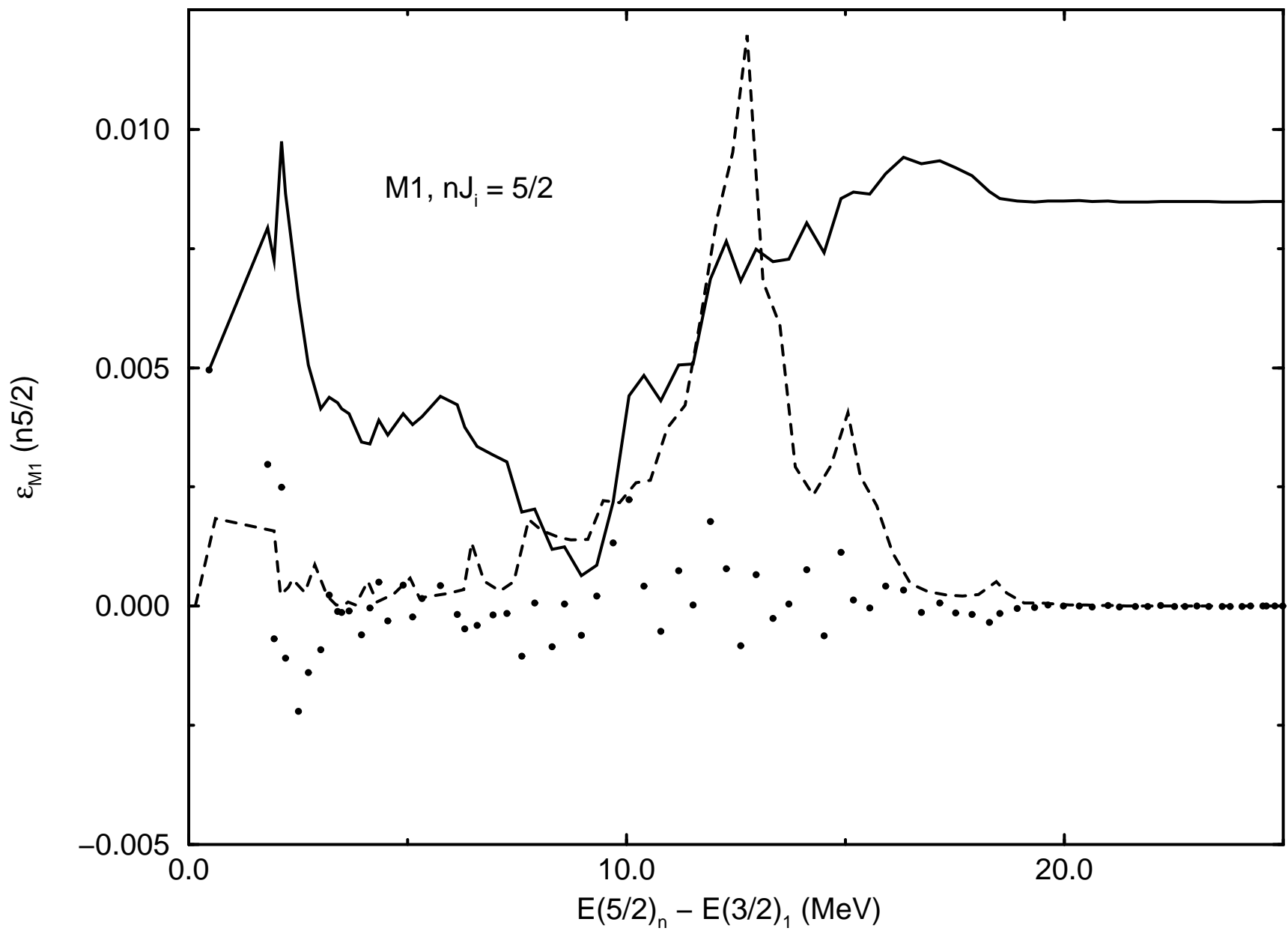
\end{document}